\documentclass{aa}  

\usepackage{graphicx}
\usepackage{gensymb}
\usepackage{array}
\usepackage{lscape}                                
\usepackage{natbib}
\usepackage{longtable}
\usepackage{color}
\usepackage{mathrsfs} 
\usepackage{subcaption}
\usepackage{etoolbox}
\usepackage[utf8]{inputenc}
\usepackage{booktabs}

\usepackage{txfonts}
\usepackage{xcolor}
\usepackage[breaklinks=true]{hyperref} 
\hypersetup{
    colorlinks=true,
    linkcolor=blue,
    filecolor=magenta,
    urlcolor=blue,
    citecolor=blue,
    }
\bibpunct{(}{)}{;}{a}{}{,} 
\begin{document}

   \title{Astrophysical parameters of LS~437 and the nature of X0726$-$260}

   \author{I.~Negueruela
          \inst{1,2}
          \and
          S.~R.~Berlanas\inst{1,3,4}
          \and
          L.~J.~Townsend\inst{5,6}
          \and
          J.~Lorenzo\inst{7}
          \and
          K.~Rubke\inst{1}
          }

   \institute{Departamento de F\'{\i}sica, Facultad de Ciencias, Universidad de Alicante, Carretera de San Vicente s/n, E03690, San Vicente del Raspeig, Spain
\and
Instituto Universitario de Investigaci\'on Inform\'atica, Universidad de Alicante, San Vicente del Raspeig, Spain
\and
Instituto de Astrof\'{\i}sica de Canarias, E-38200 La Laguna, Tenerife, Spain     
\and
Universidad de La Laguna, Dpto. Astrof\'{\i}sica, E-38206 La Laguna, Tenerife, Spain
\and
South African Astronomical Observatory, Observatory Road, Observatory 7925, Cape Town, South Africa
\and
Southern African Large Telescope, Observatory Road, Observatory 7925, Cape Town, South Africa
\and
DFISTS, EPS, Universidad de Alicante, Carretera San Vicente del Raspeig s/n,  E-03690, San Vicente del Raspeig, Spain}

   \date{Received ; accepted }
 
  \abstract
   {Be/X-ray binaries, the most common class of high-mass X-ray binaries, are characterised by OBe companions, but display a rich variety of X-ray behaviours. One of the most atypical systems is X0726$-$260, which also has the earliest optical counterpart among the whole Milky Way and Magellanic Cloud sample. }
   {We intend to improve the characterisation of the optical counterpart, LS 437, and to constrain the physical mechanisms responsible for the anomalous properties of X0726$-$260.}
   {We analyse high-quality, high-resolution optical spectroscopy of LS~437 with standard quantitative methodology to derive stellar parameters. We also make use of archival X-ray monitoring. }
   {We derive a moderate projected rotational velocity $v~\sin\,i \approx 155\:\mathrm{km\,s}^{-1}$ and a spectral type O7.5\,Ve ($T_{\mathrm{eff}}= 36\,000$~K), which makes LS~437 substantially earlier than any other Oe star in an X-ray binary. At this spectral type, the stellar wind likely contributes significantly to mass accretion, and the X-ray light curve is strongly suggestive of an orbitally modulated wind accretor. The source shows marked carbon depletion, whereas nitrogen is only slightly above solar abundance.}
   {LS~437 is the earliest Oe star known in the Galaxy, alongside HD~155806. Long-term X-ray lightcurves of X0726$-$260 strengthen the identification of a persistent 34.5~d periodicity as the orbital period, demonstrating that the X-ray emission is orbitally modulated and no X-ray outbursts have occurred over the past 30 years. Likewise, emission features in the optical spectrum indicate a remarkably stable circumstellar disk, with no sign of major structural changes over the past 40 years. All these characteristics set X0726$-$260 clearly apart from typical Be/X-ray binaries.} 


\maketitle

\section{Introduction}

Be/X-ray binaries (from now on, BeX) constitute a major class of High-Mass X-ray binaries (HMXBs) characterised by a Be star as the optical counterpart \citep[see][for a review]{reig2011}. A Be star is an early-type star that has, at some stage, displayed emission lines. Most Be stars (classical Be stars) are fast rotators surrounded by a quasi-Keplerian circumstellar disk \citep[see][for a review]{porter03}. Be stars are not far from the main sequence, generally with luminosity classes III\,--\,V \citep[cf.][]{negueruela04_lum}. In the Milky Way, they span spectral types extending from late O (O9, technically Oe) to early A (A shell stars may be a continuation at lower temperatures; cf.\ \citealt{abt97}). The fraction of Be stars is higher among early B-types (peaking at B1\,--\,B2; \citealt{zorbriot97, tarasov12}), although it may also be high around B7 (\citealt{blesson08}, but see also \citealt{mcswain05}). 

Most BeX exhibit characteristics that identify their mass donors as classical Be stars, such as double-peaked emission lines and near-infrared excesses \citep[e.g.][]{riquelme12}. An exception is the persistent X-ray source 4U~2206+54, whose counterpart, the peculiar O-type star BD~+53$^\circ$2790, displays emission lines, but characteristics atypical of a classical Be star \citep{negreig01,blay06}. The vast majority of BeX, when observed with sufficiently sensitive instruments, present X-ray pulsations, which together with power-law X-ray spectra, identify their mass accretors as magnetised neutron stars \citep[e.g.][]{boldin13,klus14,yang17}. 

Broadly speaking, BeX can be divided into two groups, according to their long-term X-ray behaviour:
\begin{itemize}
\item Transient BeX spend most of their time in quiescence, emitting X-rays below the detection threshold of survey instruments \citep[cf.][]{tsygankov17_low}. Occasionally, they undergo series of outbursts at moderate luminosities $L_{{\mathrm X}}\sim 10^{36}\:\mathrm{erg\,s}^{-1}$, generally recurring with their orbital period (Type~I outbursts) or giant ($L_{{\mathrm X}}\gtrsim 10^{37}\:\mathrm{erg\,s}^{-1}$) outbursts of longer duration (Type~II outbursts). These systems tend to have measurable eccentricities ($e\gtrsim0.3$) and orbital periods ranging from tens to a few hundred days. Their behaviour can be explained by the truncated disk model of \citet{okaneg01}. There is also a small group of BeX transients with very low eccentricity \citep[e.g.\ XTE\,J1543$-$568 or KS\,1947+300;][]{townsend11,fortin23}. 

\item Persistent BeX show much less X-ray variability. They present moderately variable low luminosity ($L_{{\mathrm X}}\lesssim 10^{35}\:\mathrm{erg\,s}^{-1}$), long ($\gtrsim 200\:$s) pulsation periods and lack of a 6.4~keV Fe line in their spectra \citep{reig1999}. \citet{lapalombara25_class} redefine the class by adding the presence of a hot blackbody component in their X-ray spectrum to their typical characteristics, and relaxing the need for long spin periods. This allows them to extend the class to objects with shorter spin periods. Nevertheless, the separation between the two groups has been somewhat diluted by the transition of the prototype of the class RX~J0440.9+4431 to an outbursting phase \citep[e.g.][]{li24}. 
\end{itemize}

With regard to their optical counterparts, all Galactic BeX with reliable spectral classifications are earlier than B2, with most presenting spectral types in a very narrow range around B0 \citep{reig2011}. A correlation is seen between the maximum historical value of the equivalent width (EW) of the H$\alpha$ emission line and the orbital period of the BeX. This is interpreted in terms of circumstellar disk truncation, since the EW of H$\alpha$ is taken as a proxy for disk size at a given time \citep{reig97,reig2011}. The correlation is loose, because other factors such as inclination to the line of sight also affect the EW \citep[see][for a careful discussion of disk variability in BeX]{reig16}. A similarly loose correlation also exists between the spin and orbital periods of BeX systems. This correlation, first noted by \citet{corbet86}, may arise from a balance between neutron star spin-up during accretion and spin-down during quiescence, but the exact mechanism is obscured by the complexity of the physical problem \citep[see][and references therein]{cheng14,xu19}.

In this context, the peculiar Be/X-ray binary X0726$-$260  (= 4U~0728$-$25) does not seem to fit any of the known classes. Discovered in the early days of X-ray astronomy, X0726$-$260 has been consistently detected at X-ray luminosities ranging from $L_{\mathrm{X}}\approx 10^{35}\:$erg\,s$^{-1}$ to $\approx6$ times brighter \citep[and references therein]{lapalombara25_0726}, with some indication of fluxes being consistently brighter during its first decade of observations. Its optical counterpart, LS~437 (= V441~Pup) was identified as a late O-type emission-line star at a relatively high distance of $\sim6$~kpc \citep{negueruela96}. Spectral classification of Oe stars is difficult, as some diagnostic \ion{He}{i} lines may present emission components \citep{negueruela04_oe}. Two approaches are possible. Insisting on the morphological character of classification, the presence of emission components is not considered, and the ratios between the composite \ion{He}{i} lines and \ion{He}{ii} lines are used as in any other O-type star. With these criteria, \citet{maiz16} classified LS~437 as O5:\,Ve. A second method to classify Oe stars is by guessing the intrinsic shape of the absorption \ion{He}{i} lines and providing a \textit{corrected} line ratio that informs more closely about the star's physical parameters. Following this approach, \citet{negueruela96} classified LS~437 as O8\,--\,9\,Ve.

The unseen companion to LS~437 is, without any doubt, a neutron star.
Using \textit{RossiXTE} observations, \citet{corbet97} found a pulsation period of 103~s, and a strong X-ray flux modulation with a 34.5~d periodicity, confirmed by subsequent observations with other satellites (and see Section~\ref{sec:fourier}), which they interpreted as the orbital period. A 2016 observation with \textit{AstroSAT} found strong pulsations at 103.144~s \citep{roy20}, while a more recent measurement with \textit{XMM-Newton} found $P_{\mathrm{spin}}=103.301$~s \citep{lapalombara25_0726}, indicating significant spin down. Its spectrum can be fitted with the typical power-law plus soft blackbody model observed in many accreting neutron stars, though alternative models provide equally acceptable fits \citep{lapalombara25_0726}.

In this paper, we present a collection of intermediate--high resolution optical spectra of the counterpart, LS~437, with the aim of deriving its stellar parameters. In combination with lower-resolution optical spectroscopy and long-term X-ray monitoring, these data allow us to constrain the nature of this peculiar system.

\section{Observations}
\label{sec:obs}

\subsection{Optical spectroscopy}
\label{sec:spec}

Several spectra of low or intermediate-low ($R\lesssim2\,500$) resolution were taken between 1999 and 2006 during observing campaigns at a number of telescopes. Details are presented in Table~\ref{tab:facilities}. These observations complement the more intensive monitoring presented in \citet{negueruela96} for the 1990\,--\,1995 interval.  

All spectroscopic data from these campaigns were reduced using the \texttt{STARLINK}\footnote{https://starlink.eao.hawaii.edu/starlink/WelcomePage} software suite \citep{currie14}, including \texttt{ccdpack}\footnote{https://starlink.eao.hawaii.edu/devdocs/sun139.htx/sun139.html}, \texttt{figaro} \citep{shortridge14}, and \texttt{dipso} \citep{howarth14}. This historical dataset is highly heterogeneous, comprising diverse instrumental configurations and resolutions.

\begin{table}[h!]
\caption{Summary of low-resolution spectra of LS~437. \label{tab:facilities}}
\centering
\begin{tabular}{lccc}
\hline\hline
\noalign{\smallskip}
Observatory & Dates & Instrument & Range (\AA) \\
\noalign{\smallskip}
\hline
\noalign{\smallskip}
ESO 1.5~m & 1999 Nov 2 & B\&C & 3600\,--\,5600 \\
ESO 1.5~m &2000  Sep 14 & B\&C & 3600\,--\,5600 \\
NOT & 2004 Oct 3 & ALFOSC & 3800\--\,5100\\
NOT & 2004 Oct 3 & ALFOSC & 3800\--\,6800\\
NOT & 2005 Oct 4 & ALFOSC & 3800\--\,5100\\
SAAO 1.9m & 2006 May 6 & Spect & 6200\,--\,7400\\
SAAO 1.9m & 2006 May 7 & Spect & 6200\,--\,9000\\
SAAO 1.9m & 2006 May 8 & Spect & 3800\,--\,5600\\
\noalign{\smallskip}
\hline
\end{tabular}
\end{table}

Additionally, a number of high-intermediate ($R>5\,000$) resolution spectra of LS~437 have been taken throughout the years and were used for spectroscopic analysis. The earliest spectrum was taken with the ESO Multi-Mode
Instrument (EMMI) on the New Technology Telescope (NTT) at La Silla Observatory (Chile) on the night of  2006 February 16. A general description of the observing run is given in \citet{marco09}. The red arm of the instrument was used in echelle mode. The detector consisted of a  mosaic formed by two thin, back-illuminated and AR-coated  $2048\times4096$ MIT/LL CCDs with $15\,\mathrm{\mu m}\times15\,\mathrm{\mu m}$ pixels, mounted side-by-side, with an 47-pixel overscan gap in between (in the x direction). We used the echelle grating \#9 cross-dispersed with grism \#3 on the red arm. This configuration provides $R\approx10\,000$ over the 4000\,--\,7900\,\AA\ range, with a small gap around 4950\,\AA.

\begin{figure*}[t!]
   \centering
   \includegraphics[width=\textwidth]{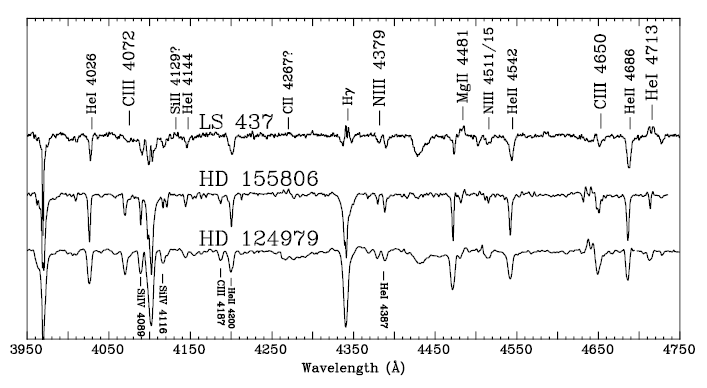}
      \caption{Classification spectrum of LS~437 together with two reference stars, the mild Be HD 155806 (O7.5\,Ve) and the moderate rotator HD~124979 (O7.5\,IV). Spectral lines of special interest are marked in larger font. See main text for a detailed description.}
         \label{fig:bluespec}
   \end{figure*}

Due to the low efficiency of the red arm in the blue part of the spectrum, we complemented these observations with grating \#3 in the blue arm. The detector in the blue arm was a $1024\times1024$ Textronik CCD with $24\,\mathrm{\mu m}\times24\,\mathrm{\mu m}$ pixels. This configuration gave $R=3\,400$ over the 3925\,--\,4380\,\AA\ range. We also used this grating with a redder central wavelength to obtain a spectrum covering 4380\,--\,4800\,\AA. The concatenation of these two spectra is shown in Fig.~\ref{fig:bluespec}.

We obtained several optical spectra with the Southern African Large Telescope (SALT) starting in October 2020, using the High Resolution Spectrograph (HRS) in either medium or low resolution modes. Among these, we selected the spectrum taken in low-resolution mode on 2022 March 27, which has the highest signal-to-noise (SNR) ratio in the whole dataset. This spectrum has resolving power $R\approx16\,000$, and was reduced using the telescope primary reduction pipeline and the MIDAS package of \citet{kniazev16}. The blue arm covers the range 3750\,--\,5400\,\AA, although the spectrum is only useful from $\approx 4000$\,\AA, due to very low SNR and poor order tracing. The red arm covers from 5450\,\AA\ to 8790\,\AA, although the SNR is low shortwards of $\approx 5600$\,AA. 

Finally, we observed LS~437 with the Gran Telescopio Canarias (GTC) using the integral fibre unite mode (IFU) of the Multi-Espectrógrafo en GTC de Alta Resolución para Astronomía (MEGARA) spectrograph. The IFU mode covers a field of view (FoV) of 2.5$\times$11.3~arcsec, complemented by eight additional 7-fiber minibundles placed on the outer part of the FoV for sky subtraction.  We obtained several spectra in the mid-resolution configuration of MEGARA using four different volume phase holographic gratings (VPHs). On the night 2019 December 4, two spectra were taken using the MR-R and MR-U setups. These configurations gave $R=12\,000$ over the 6243\,--\,6865\,\AA\ and 3920\,--\,4282\,\AA\ ranges, respectively. On the nights 2020 February 8 and 9, we obtained two spectra using the MR-UB setup and two others using the MR-B setup. These configurations provided similar spectral power ($R=12\,000$) over the  4226\,--\,4625\,\AA\ and 4585\,--\,5025\,\AA\ ranges, respectively. The spectra were reduced using the MEGARA Data Reduction Pipeline \citep[DRP, see][]{cardiel18,pascual18,pascual19}, and combined for analysis into a single spectrum.

\subsection{X-ray data}

We downloaded long-term lightcurves from the archives of all missions that have provided long-term X-ray monitoring. The oldest dataset is that of \textit{RossiRXTE} (2\,--\,12~keV), which covers from MJD~50\,087 to 55\,926. A similar energy range (2\,--\,20~keV) is provided by \textit{MAXI} for MJD between 55\,066 and 59\,997, providing thus some overlap. The \textit{Neil Gehrels Swift} observatory monitors a harder energy band (15\,--\,50~keV). The observations analysed here span the MJD~53\,416\,--\,59\,996 range.

\section{Analysis}
\subsection{Spectrum description}
\label{sec:descrip}

\begin{figure}[t!]
   \centering
   \includegraphics[width=0.8\columnwidth]{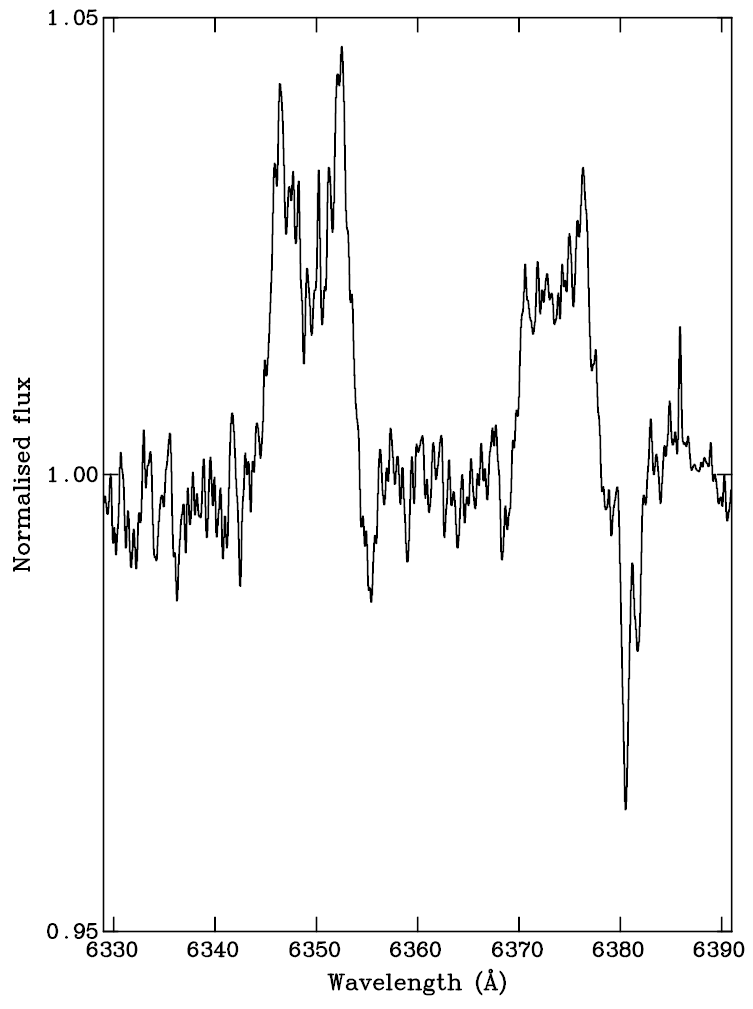}
      \caption{A detail of the SALT/HRS spectrum, displaying the region comprising the \ion{Si}{ii}~6347\,\AA\ and \ion{Fe}{ii}~6371\,\AA\ emission lines.}
         \label{fig:specbits}
   \end{figure}

The broadband optical spectrum of LS~437 is dominated by moderately strong emission lines of \ion{H}{i} and \ion{He}{i}. Weaker emission features of \ion{Fe}{ii} and other metals are visible in the yellow and red ($\approx$ 5\,000 to 7\,500\,\AA) regions observed in the EMMI echelle and SALT/HRS spectra. An example is shown in Fig.~\ref{fig:specbits}. The only prominent photospheric lines correspond to \ion{He}{ii}, clearly identifying the source as an O-type star.

The classification spectrum of LS~437 observed with EMMI is displayed in Fig.~\ref{fig:bluespec}, together with two suitable comparison stars. HD~155806 is the only catalogued Be star earlier than O8 in the Milky Way. Its Be characteristics are weak \citep[see][]{negueruela04_oe}, and here mostly manifested through the emission wings in \ion{He}{i}~4713\,\AA\ and the asymmetry of H$\gamma$, due to a weak emission component. In contrast, LS~437 presents H$\gamma$ almost completely filled in and a double peak rising above the continuum. HD~124979 was classified as O7.5\,IV(n)((f)) by \citet{sota14}. Its rotational velocity has been recently estimated at $\approx 260\:\mathrm{km\,s}^{-1}$ \citep{holgado22}, and thus its spectrum provides a closer match to LS~437, since HD~155806 shows narrow lines -- its projected rotational velocity is around $\approx 50\:\mathrm{km\,s}^{-1}$ \citep{simondiaz17}, and thus the line profiles are dominated by instrumental broadening in the $R\approx5\,000$ spectrum shown in Fig.~\ref{fig:bluespec}, taken from \citet{negueruela04_oe}.

One obvious difference between LS~437 and the comparison stars is the weakness of \ion{He}{i}~4471\,\AA, due to emission infilling. Emission in \ion{He}{i} lines such as $\lambda$6678 or $\lambda$5875 is common in early Be stars, but \ion{He}{i}~4471\,\AA\ remains purely photospheric even in some O9e stars \citep{negueruela04_oe}. It is seen in emission in the Oe star HD~39680 \citep{negueruela04_oe}, and in several Oe stars recently discovered in the SMC \citep{golden16}, most likely reflecting a higher temperature of the decretion disk when compared to Be stars. Since the ratio \ion{He}{i}~4471\,\AA/\ion{He}{ii}~4541\,\AA\ serves as the primary criterion for spectral classification of O-type stars, a fully meaningful spectral type cannot be given for these Oe stars, as discussed in the Introduction.

Another indication of a high disk temperature is the presence of \ion{Mg}{ii}~4481\AA\ in emission. To our knowledge, this emission feature has not been reported in any other Oe star, although it is very likely filled in in the spectrum of HD~39680 presented by \citet{negueruela04_oe} and is strong in more recent spectra of this object present in the IACOB database \citep{simon11,simon15}. Both HD~155806 and HD~39680 display double-peaked emission in \ion{C}{ii}~4267\,\AA, another line not seen in Be stars. A very weak double-peaked emission feature may be guessed in the spectrum of LS~437, but its presence is doubtful. As seen in Fig.~\ref{fig:specbits}, the \ion{Si}{ii}~6437\,\AA\ line is strongly in emission. Some of the lower resolution spectra strongly suggest that the \ion{Si}{ii}~4129\,\AA\ doublet may also be weakly in emission, although this is not obvious in the high-resolution spectra because of the complexity of the spectral region and also because this \ion{Si}{ii} feature is a doublet. Note that photospheric absorption is not expected for either \ion{Si}{ii} or \ion{C}{ii} at these early types \citep{sota14}.

The most striking difference between LS~437 and the comparison stars is the weakness of its \ion{C}{iii} lines. Both $\lambda$4072 and $\lambda$4650 are much weaker than in normal mid- or late-O stars, while $\lambda$4187 cannot even be detected. This pronounced depletion of carbon is not accompanied by an obvious nitrogen enhancement, as the \ion{N}{iii} lines in the spectrum of LS~437 are directly comparable to those in HD~124979. The carbon deficiency probably accounts for the absence of an obvious \ion{C}{ii}~4267\,\AA\ emission feature, as seen in the other Oe stars. An alternative explanation would be emission infilling of the \ion{C}{iii} lines, as there may be some variability in their strength (although this is very dependent on the normalisation of the low-resolution spectra). Nevertheless, it is difficult to imagine a decretion disk hot enough to drive \ion{C}{iii} into emission.

\subsection{Spectral type and distance}
\label{sec:type}

Although we cannot use the main criterion for spectral classification, there are several secondary criteria that can be used to constrain the spectral type of LS~437. The ratios \ion{He}{i}~4144\,\AA/\ion{He}{ii}~4200\,\AA\ and \ion{He}{i}~4387\,\AA/\ion{He}{ii}~4542\,\AA\ unambiguously place the spectral type close to O7 \citep{sota11}. The \ion{He}{i} lines become very weak at earlier types, with $\lambda$4144 nearly undetectable by O6. Conversely, at O8 they are comparable in strength to the \ion{He}{ii} lines, though still weaker. As can be seen in Fig.~\ref{fig:bluespec}, these ratios in LS~437 closely match those of the two O7.5 stars shown, and we therefore adopt this as the most likely spectral type for the object. It is worth noting that, although the spectrum presented in \citet{negueruela96} is noisy, the ratio \ion{He}{i}~4471\,\AA/\ion{He}{ii}~4541\,\AA\ approximately corresponds to O7.5, strongly suggesting that \ion{He}{i}~4471\,\AA\ was purely photospheric at the time. Consistently, the EW of H$\beta$ that they report is smaller than in any modern spectrum. Assuming that the \ion{He}{i} line was still in-filled, the authors favoured a slightly later type.

The strength of \ion{He}{ii}~4686\,\AA\ relative to the other \ion{He}{ii} lines clearly identifies LS~437 as a dwarf. This line exhibits a very strong negative luminosity dependence. Already in the spectrum of HD~124979 (luminosity class IV), it is considerably weaker.

The \textit{Gaia} DR3 parallax for LS~437 is $\varpi=0.096\pm0.017$~mas. Using the model of \citet{bailer23}, this translates into a photogeometric distance of $7.6^{+1.0}_{-0.8}$~kpc. The geometric distance is 7.9~kpc, with larger uncertainties. From a careful analysis of the SED, \citet{riquelme12} calculated $7.0\pm0.5$~kpc, although assuming a slightly later O8.5\,V spectral type. In view of these values, we will adopt 7.6~kpc as the distance to LS~437.

\subsection{Stellar parameters}
\label{sec:params}

   \begin{figure*}[t]
   \centering
   \includegraphics[width=13cm]{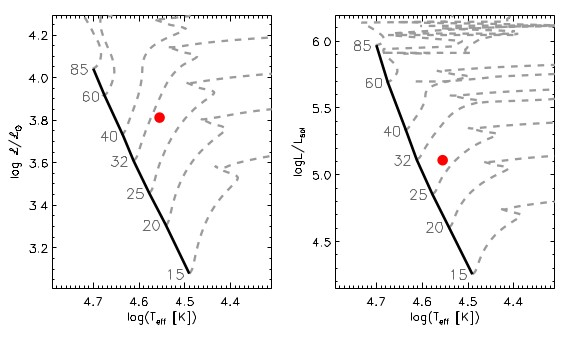}
      \caption{The HR diagram (right panel) and its spectroscopic version (left panel) using Geneva rotating evolutionary stellar tracks ($v_{\mathrm{ini}}/v_{\mathrm{crit}}\sim$0.4). We note that the \( \mathcal{L} \) parameter is defined in terms of effective temperature and surface gravity as $T^{4}_{\mathrm{eff}}$ /$g$, which is equivalent to the  $L$/$M$ ratio. In both panels, the red dot identifies the position of LS~437. Evolutionary tracks are labelled with the initial mass of the star.}
         \label{hrd}
   \end{figure*}

We determined the rotational and main stellar parameters for LS~437 by performing a quantitative spectroscopic analysis based on synthetic FASTWIND models \citep{santo1997, puls2005}, as used in \citet{berlanas25} or \citet{herrero22}. For this purpose, we used the higher resolution spectra from MEGARA, EMMI and SALT/HRS.

To derive the rotational parameters, we used the \texttt{iacob-broad} tool \citep{ssimon07,ssimon14}.
This is a user-friendly IDL procedure for the characterization of line-broadening in OB stars, based on the combination of  Fourier Transform (FT)  and goodness-of-fit (GOF) methodologies. It allows the user to  simultaneously determine the stellar projected rotational velocity ($v\sin\,i$) and the amount of extra broadening (assuming a radial-tangential profile, $v_{\mathrm{mac}}$) from a specifically selected diagnostic line. The FT technique identifies the first zero in the Fourier transform of a given line profile \citep{gray08, ssimon07}, while the GOF method  compares the observed profiles with a synthetic line profile that is convolved with different values of $v\sin\,i$ and $v_{\mathrm{mac}}$ to obtain the best-fit by means of a $\chi^{2}$ optimization. 

Finding a line in the spectrum of LS~437 suitable for rotational velocity determination is challenging. The traditional diagnostic feature for O-type stars, \ion{O}{iii}~5591\,\AA, is not included in the range covered by the MEGARA spectrum, and cannot be clearly seen in the EMMI or SALT/HRS spectra. The \ion{Si}{iv} lines lie in the region with lower SNR and are blended into the wings of H$\delta$. Lacking any usable metallic line, we based the  analysis on the \ion{He}{i}~4387\,\AA\ line. Although it is weak, it does not seem contaminated by emission and does not suffer from strong Stark broadening. We derived consistent rotational parameters from all available spectra, obtaining $v\sin\,i = 155\: \mathrm{km\,s}^{-1}$ and macroturbulent broadening $v_{\mathrm{mac}} = 145\:\mathrm{km\,s}^{-1}$, with excellent agreement between values derived from both the FT and GOF techniques. Uncertainties are in the range 10\,--\,20$\%$.

Once the rotational velocity parameters were established, we used them as input for the \texttt{iacob-gbat} tool \citep{ssimon11} to derive the main spectroscopic parameters, i.e. effective temperature ($T_{\mathrm{eff}}$), surface gravity ($\log\,g$) and helium abundance ($Y(\mathrm{He})$), defined as $N(\mathrm{He})/N(\mathrm{H})$. This tool compares the observed and synthetic line profiles (from FASTWIND models, in our case) by applying a $\chi^{2}$ algorithm. It computes the line-by-line $\chi^{2}$ distributions, estimating the goodness-of-fit for each model within a subgrid of models selected from the global grid. Then it iteratively computes the global $\chi^{2}$ distribution, from which the final parameter values and their associated uncertainties are estimated. The final parameters correspond to the mean values computed from all the models located within the 1-$\sigma$ confidence level of the total $\chi^{2}$ distributions (after each model has been weighed by its corresponding $\chi^{2}$ value). Then their uncertainties are given by the standard deviation within the 1-$\sigma$ level. Our grid of models covers the wide range of stellar and wind parameters considered for standard OB-type stars. 

We discarded all H and He lines with evident emission features and based the analysis on H${\delta}$,  \ion{He}{i}~$\lambda$4143, $\lambda$4387 and \ion{He}{ii}~$\lambda$4200, $\lambda$4541, $\lambda$4686. Each spectrum was analysed independently, but all yield very similar temperatures, always in the range 35\,--\,36~kK (including estimated uncertainties). We make note that the emission present in the centre of the H ${\delta}$ line was cleaned to leave the line wings isolated. Nevertheless, this was insufficient to constrain precisely the surface gravity. Accordingly, we adopted $\log\,g$ as 3.8~dex, within the range of possible values, by taking into account the spectral type and the strength of \ion{He}{ii}~4686\,\AA, which suggests that the star is not close to the ZAMS. 

In all our spectra, a change of $+0.1$~dex in the gravity (i.e. the range of possible values for a luminosity class V star not on the ZAMS) changes the derived temperature by less than 500~K, which lies within the typical 1~kK uncertainty, and thus does not affect the final estimates. From the best fitting models, we obtained a mean value for $T_{\mathrm{eff}}$  of $35.9\pm 1.0$~kK. These $T_{\mathrm{eff}}$  and $\log\,g$ values were later used to determine the carbon and nitrogen abundances (see next subsection).

Assuming a distance of $\sim7.6$~kpc, the \texttt{iacob-gbat} tool also provides the stellar radius ($R$), luminosity ($\log (L/\mathrm{L}_{\odot})$) and spectroscopic mass ($M_{\mathrm{sp}}$) when the absolute magnitude, $M_{V}$, is specified. For this, we adopt the reddening determination of \citet{riquelme12} and the \cite{rieke85} extinction law, finding A$_{V}$ = 2.1 and thus $M_{V} = -4.70$. The main spectroscopic parameters are summarized in Table~\ref{tab:params}. The traditional Hertzsprung-Russell diagram (HRD) and its spectroscopic version \citep[sHRD, see][]{langer14} are shown in Fig.~\ref{hrd}, together with Geneva evolutionary tracks \citep{ekstrom12}. The position of LS~437 in both diagrams (marked with a red dot) is consistent with an evolutionary mass close to $30\:\mathrm{M}_{\odot}$, confirming that the adopted surface gravity is appropriate. We find $M_{\mathrm{sp}}$/$M_{\mathrm{ev}} <1.0$, noting that the spectroscopic mass has been derived using the gravity corrected from centrifugal acceleration.

The discrepancy between evolutionary and spectroscopic masses in O-type stars, commonly known as the mass-discrepancy problem, has been widely discussed in the literature \citep[see, e.g.][]{herrero92,massey05, martins12, mahy15, mahy20b, berlanas25}. However, no firm consensus has yet been reached regarding its origin. Spectroscopic masses are very sensitive to the adopted extinction, which is subject to significant and often poorly constrained uncertainties, particularly related to the choice of extinction law. On the other hand, evolutionary masses may change significantly depending on the evolutionary models employed. Since the discrepancy for LS~437 is entirely typical of O-type stars, we cannot draw any conclusions.

Given that the analysis of spectra spanning close to 15 years yields mutually consistent values for $T_{\mathrm{eff}}$, we are reassured that variable emission components do not contribute in any significant way to the diagnostic lines used. Our value for $T_{\mathrm{eff}}$ is in excellent agreement with the derived spectral type, according to modern calibrations \citep{martins05,holgado25}. The inferred absolute magnitude and luminosity are also consistent with expectations, although somewhat higher than average, which is in line with our adoption of $\log\,g=3.8$, based on the intensity of \ion{He}{ii}~4686\,\AA. The resulting spectroscopic mass is fully consistent with the average value found by \citet{holgado25} for O7.5\,V stars. The evolutionary mass tabulated for this type is $24\:\mathrm{M}_{\odot}$  \citep{martins05}.
   
   \begin{table}[t]
   \centering
   \caption{Derived spectroscopic parameters for LS~437.}
         \label{tab:params}
         \begin{tabular}{l c}
            \hline
            \noalign{\smallskip}
            {Parameter}   & {Value} \\
            \hline
            \noalign{\smallskip}            
            {$v~\sin\,i$ [km s$^{-1}$]}      &  $155\pm25$ \\            
            {$v_{\mathrm{mac}}$ [km s$^{-1}$]}      &   $145\pm25$\\     
            {$T_{\mathrm{eff}}$ [kK]}      &  35.9 $\pm$ 1.0 \\
            {$\log\,g$   [dex]}  &  $3.8\pm0.1^{\:,a}$ \\
            {$A_{V}$  [mag]}  &  2.3$^{\:,b}$ \\ 
            {M$_{V}$  [mag] }  &  $-4.7^{+0.2}_{-0.3}$$^{\:,c}$ \\
            {$R$  [R$_{\odot}$]}   &  $9.6 \pm0.2$$^{\:,d}$ \\
           {log\,($L$/L$_{\odot}$) [dex]} &  5.14 $\pm$ 0.03$^{\:,d}$ \\
            {$M_{\mathrm{sp}}$ [M$_{\odot}$]} &  22.2 $\pm$ 0.5$^{\:,d}$ \\
            \noalign{\smallskip}
            \hline
         \end{tabular}
         \begin{minipage}{\columnwidth} 
\vspace*{0.1cm}
\footnotesize
$^a$ Adopted. Errors represent values consistent with $T_{\mathrm{eff}}$ determination, but lower $\log\,g$ already corresponds to luminosity class III.\\
$^b$ Adopted from \citet{riquelme12}. Uncertainties depend on the deviation from the standard extinction law, which cannot be quantified.\\
$^c$ Formal error due to uncertainty in distance. The dependence on $A_V$ cannot be quantified.\\
$^d$ Formal errors from the procedure. Dependence on distance ignored.
\end{minipage}
   \end{table}

\subsection{Abundances}
\label{sec:abunds}

To derive carbon (C) and nitrogen (N) abundances, we use a new grid of HHeCNO models calculated with version 10.6.5 of the FASTWIND code and the distributed computation system HTCondor\footnote{http://research.cs.wisc.edu/htcondor/: the supercomputer
facility at Instituto de Astrofisica de Canarias.}. These models include the improved N model atom developed and described in \cite{rivero11}.

Given the weakness of diagnostic lines in the spectrum of LS~437 and the relatively low S/N in the blue-violet region, we only aim to constrain abundances and estimate reliable limits through supervised synthetic profile matching. This approach was preferred over an automated $\chi^{2}$ minimization, as the latter can be heavily biased by pixel-to-pixel noise fluctuations and small continuum placement uncertainties in a noise-dominated regime. Starting from solar values, we adjusted the abundances iteratively to achieve a better match with the observations. The adopted rotational velocity, $T_{\mathrm{eff}}$, $\log\,g$  and $Y$(He) were those derived from the \texttt{iacob-gbat} analysis. Oxygen cannot be constrained due to the lack of detectable lines and  still imprecise atomic models used in current atmospheric models. Therefore, we focus the analysis on C and N abundances. We have used all lines available in our spectra, except for those too faint or blended with other lines.

The best fitting model for our spectra, assuming $T_{\mathrm{eff}}$= 36.0~kK, $\log\,g$  = 3.8~dex and $Y$(He)= 0.06~dex, corresponds to carbon and nitrogen abundances of 8.05~dex and 8.10~dex, respectively, where abundances are defined as $\epsilon_{X}$ = 12 + log(X/H), with solar reference values of 8.35 (C) and 7.80 (N) from \citet{nieva11}. Assuming the derived uncertainties in temperature, we checked the effect of varying $T_{\mathrm{eff}}$ by 1000~K (taking into account the possible corresponding gravity variation). We derive carbon and nitrogen abundance values of 7.75~dex and 7.90~dex, respectively, when assuming $T_{\mathrm{eff}}$= 35~kK. The difference in abundance between these models gives us the uncertainties of our estimates. To ensure the reliability of these constraints, we performed a sensitivity analysis by overplotting synthetic spectra with variations of $\pm 0.05$ dex, confirming that such differences are detectable. The best fitting models are shown in Fig.~\ref{bfm}.

We also examined the possibility that the continuum flux includes a contribution of continuum emission from the circumstellar disk. Assuming a factor of a 10 percent \citep[see][]{riquelme12}, we find no variations in the derived stellar parameters (derived temperature of $36.0\pm 0.8\:$kK). Even when considering a higher contribution from the disk to the total flux of 20\,\%, the change in temperature is still within the derived uncertainties ($T_{\mathrm{eff}} =  35.0 \pm 1.1\:$kK). We thus conclude that the possible circumstellar contribution to the stellar flux has a negligible effect on the derived stellar parameters.

\subsection{Disk properties}
\label{sec:disk}
\begin{figure}
   \centering
   \includegraphics[width=0.493\columnwidth]{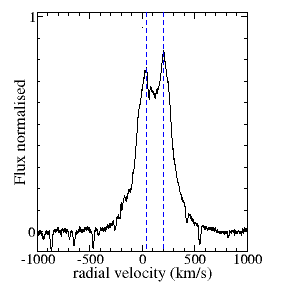}
    \includegraphics[width=0.493\columnwidth]{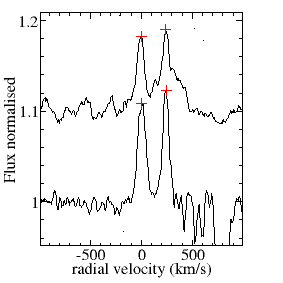}
      \caption{Emission lines seen in the spectrum of LS~437, taken on 2022 March 27, in velocity space. The left panel shows H$\alpha$, the strongest emission line in the spectrum, with the two lines marking the position of the peaks. The right panel shows two \ion{He}{i} lines, $\lambda$5015 (top) and $\lambda$5875 (bottom), with crosses marking the position of the peaks. The \ion{Fe}{ii}~$\lambda$5018 line is blended into the red wing of $\lambda$5015. }
         \label{fig:lines}
   \end{figure}

The contribution of the circumstellar disk is evident in the emission lines.
All of them appear double-peaked in all the high-resolution spectra. Representative line profiles are shown in Fig.~\ref{fig:lines}. The strong H$\alpha$ line has a typical morphology for a Be star. According to the classical shape classification of \citet{hanuschik96}, its profile corresponds to a disk seen at a moderately high inclination, as it is intermediate between the examples illustrated in their fig.~2. It displays no signs of the winebottle-type inflections that characterise lower inclination systems. Such features are typically found for inclinations lower than $i=30\degr$ to $45\degree$, depending on the physical properties of the disk \citep{hummel94,lailey24}. The \ion{He}{i} lines, which have profiles more typical of optically thin features, are consistent with this inclination estimate.

%

Interestingly, the equivalent width (EW) of H$\alpha$ measured on our spectra ranges from $-7$ to $-10\,\AA$. Such values are comparable to the highest values measured by \citet{negueruela96} in the 1990s. 

We made use of the observed emission lines in an attempt to constrain the properties of the circumstellar disk around LS~437. Since the line profiles are quite similar in all three high-resolution spectra, we selected the SALT/HRS spectrum for this purpose, as it has the broadest spectral coverage and higher SNR over most of the range. We normalised the spectrum and smoothed it with a Gaussian filter ($\sigma=0.2\:$\AA) to improve the determination of peaks and wing edge detection. We measured the characteristics of the most prominent emission lines, which are reported in Table~\ref{tab:disksize}. The parameter $v_{\mathrm{peak}}$ is defined as half the velocity separation between the two peaks, while $v_{\mathrm{wings}}$ corresponds to half the separation between the points where the line wings meet the continuum. For the \ion{H}{i} and \ion{He}{i} lines, we made no attempt to subtract the underlying absorption profiles. Only lines for which $v_{\mathrm{wings}}$ can be reliably measured are included in Table~\ref{tab:disksize}. Many other weaker lines do not allow a good definition of this parameter. The strong \ion{He}{i}~5015\,\AA\ is not included because it is blended with \ion{Fe}{ii} (see Fig.~\ref{fig:lines}) and \ion{He}{i}~6678\,\AA\ is excluded because of blending with the absorption feature due to \ion{He}{ii}~6683\,\AA.

The peak separation may be used for a rough estimation of disk size, since they are expected to reflect the outermost region from which emission in a given ion is arising. Following \citet{huang72}, we have:

\begin{equation}
\frac{R_{\textrm{out}}}{R_{*}} = \left( \frac{v \sin\,i}{v_{\textrm{peak}}} \right)^2\, ,
\end{equation}

where $R_{\mathrm{out}}$ estimates the radial distance from the central star at which this outermost region lies.
Taking $R_{*}= 9.6\:\textrm{R}_{\sun}$ and $v\sin\,i = 155\:\textrm{km\,s}^{-1}$ from our spectroscopic solution and the peak separations measured, we find the values for $R_{\textrm{out}}$ that are listed in Table~\ref{tab:disksize}. As expected, $R_{\textrm{out}}$ correlates rather well with the transition energy of each line. The \ion{He}{i} lines and H$\beta$ all give similar values around $15\:R_{*}$, with the metallic lines giving slightly shorter distances. The value for H$\alpha$ is almost certainly unphysical, as it is optically thick, and other physical processes can determine the position of the peaks \citep{hummel94}.

\begin{table}
\caption{Measured disk parameters and derived disk sizes for different emission lines. Measured values are projected, and should be divided by $\sin\,i$ to yield actual physical magnitudes. All velocity measurements have an error of $\pm3\:\mathrm{km\,s}^{-1}$.\label{tab:disksize}}      
\centering
\begin{tabular}{l c c c}        
\hline\hline 
\noalign{\smallskip}
Spectral line & $v_{\mathrm{peak}}$ & $v_{\mathrm{wings}}$ & $R_{\mathrm{out}}$ \\  
 & (km\,s$^{-1}$) & (km\,s$^{-1}$) & ($R_{*}$) \\
\noalign{\smallskip}
\hline
\noalign{\smallskip}
Mg\,\textsc{ii}~4481\,\AA& 155 & 233 & $9.4\pm0.6$\\
He\,\textsc{i}~4713\,\AA & 132 & 294 & $12.9\pm0.8$\\
H$\beta$                   & 120 & 367 & $15.6\pm0.9$\\
He\,\textsc{i}~5876\,\AA & 126 & 251 & $14.2\pm0.9$\\
Si\,\textsc{ii}~6347\,\AA& 141 & 226 & $11.3\pm0.7$\\
H$\alpha$                  &  82 & 611 & $33.4\pm2.0$\\
He\,\textsc{i}~7065\,\AA & 118 & 221 & $16.1\pm1.0$\\
\noalign{\smallskip}
\hline                        
\end{tabular}
\end{table}

\subsection{X-ray analysis}
\label{sec:fourier}

\begin{figure*}
   \centering
    \includegraphics[width=0.48\textwidth]{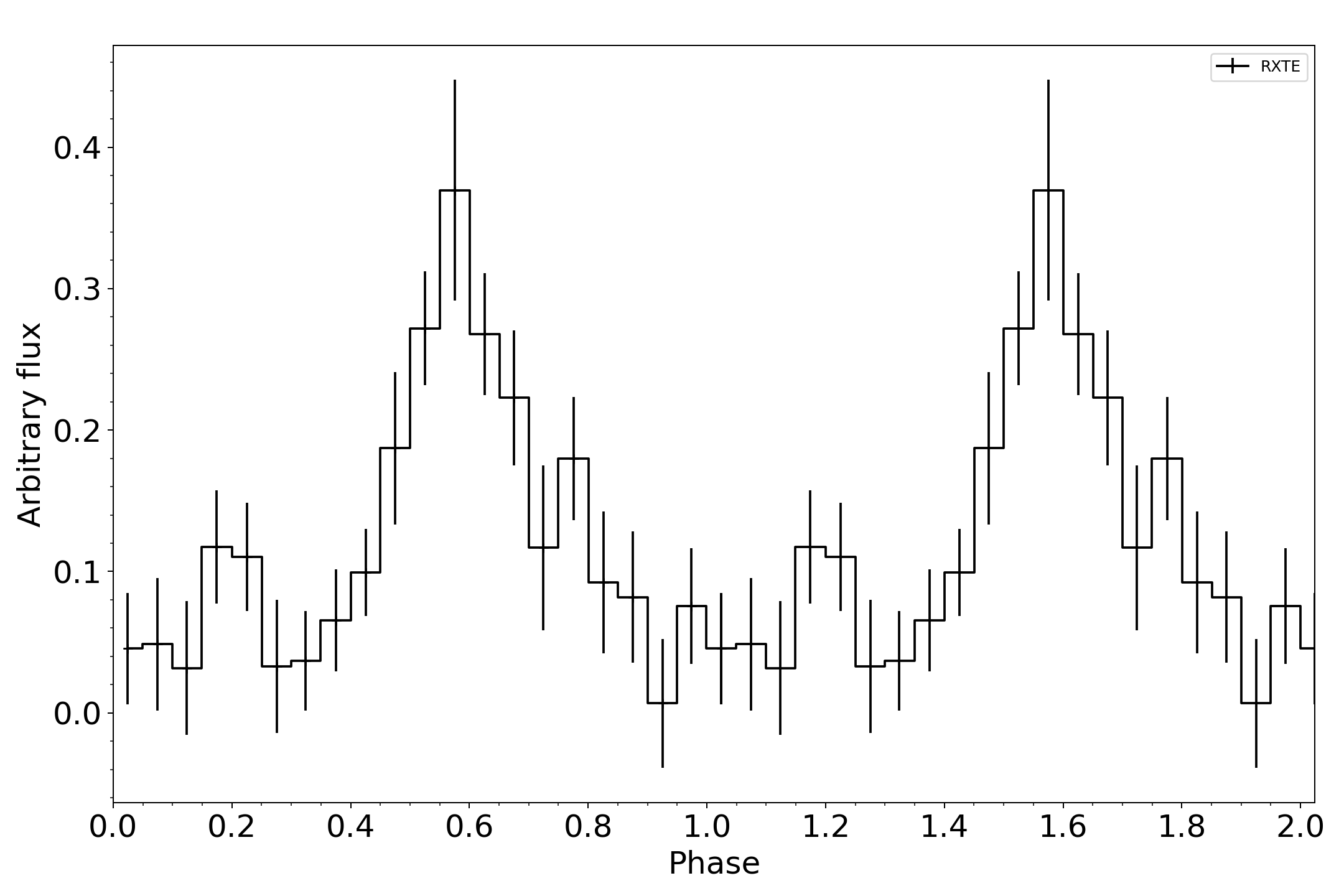}
    \includegraphics[width=0.48\textwidth]{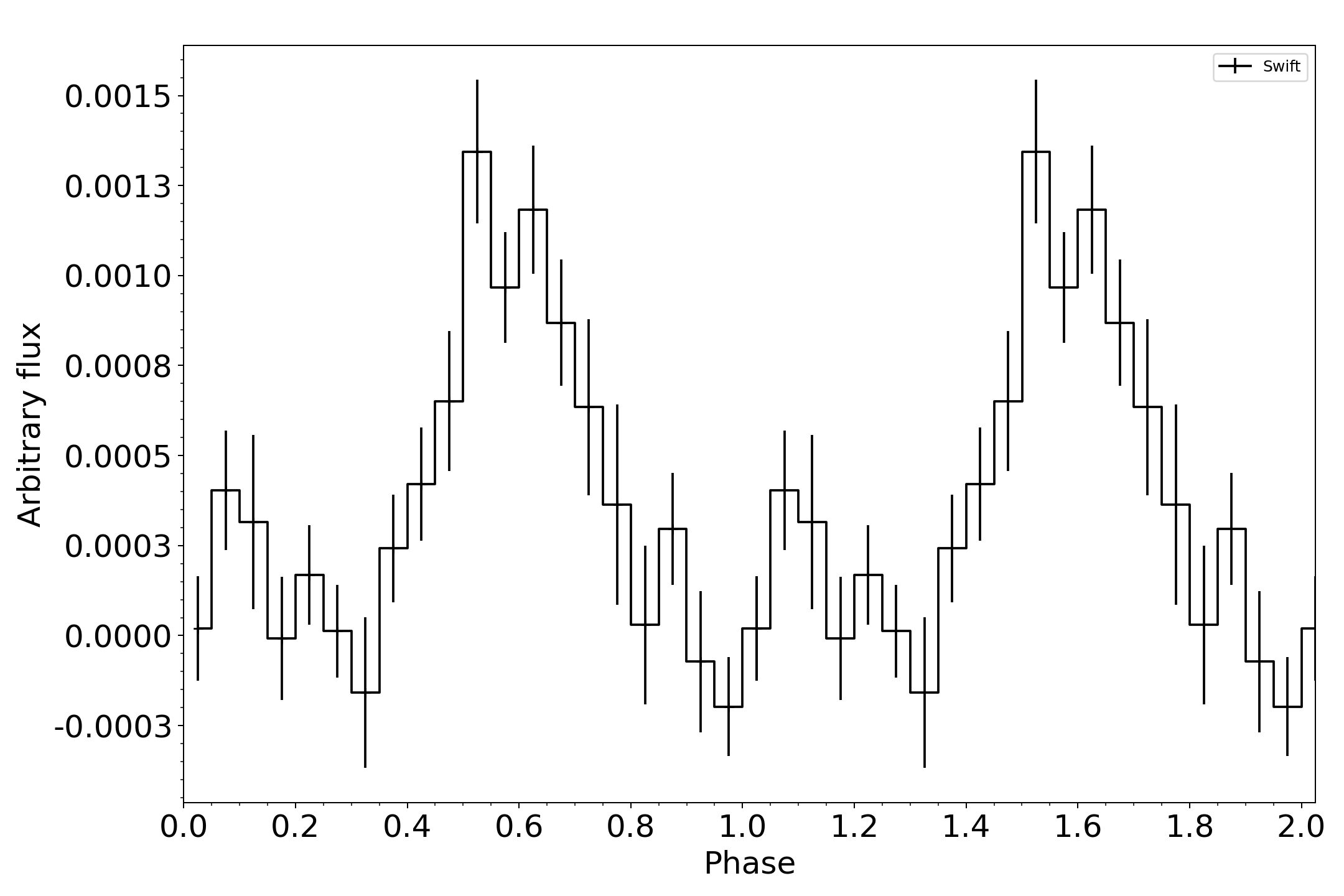}
   \caption{Long-term X-ray lightcurves folded on the period derived from FFT analysis. The left panel shows the \textit{RossiRXTE} (2\,--\,12~keV) lightcurve extending from Jan 1996 to Dec 2011. The right panel displays the \textit{Neil Gehrels Swift} observatory (15\,--\,50~keV) lightcurve obtained between Feb 2005 and Feb 2023. For ease of comparison, in both plots $T_{0} =$ 50\,000 (MJD) has been chosen.} 
              \label{fig:folds}%
    \end{figure*}

Running a Lomb-Scargle periodogram on the full \textit{RXTE}/ASM light curve reveals a highly significant period at $34.541\pm0.007$~d, which is fully consistent with the $34.46\pm0.12$~d found by \citet{corbet97} for the first 1.5~years of data, which they interpreted as the orbital period of the system. Folding the light curve on this period results in the plot shown in Fig.~\ref{fig:folds} (left panel). Here we use phase bins of 0.05 and plot two phase cycles for clarity. We arbitrarily choose MJD 50\,000 as phase zero to facilitate comparison with other missions discussed below. The shape is almost identical to that shown by \citet{corbet97}, based on a much shorter timespan. There is a hint of a secondary peak around phase 0.2, which is not exactly half an orbital period away from the primary peak at $\phi \approx 0.6$.

We checked the \textit{Swift}/BAT and \textit{MAXI} long-term light curves to try and confirm whether this secondary peak is real. We found highly significant periods in the power spectra of the \textit{Swift} and \textit{MAXI} light curves at $34.534\pm0.006$~d and $34.534\pm0.008$~d, respectively. The two periods are identical within their formal errors (indeed the calculated values differ by only 7~ms), and also consistent within errors with the \textit{RXTE} value. \citet{corbet16} had reported a value $34.548\pm0.010$~d using about 12 years of \textit{Swift} data, which is again consistent within the errors. The folded \textit{Swift} light curve is shown in the right panel of Fig.~\ref{fig:folds}. To allow better comparison, we also used MJD 50\,000 as phase zero. The \textit{MAXI} light curve is not shown, as it is almost identical. The position of the main peak is at approximately the same phase in both plots, although it is somewhat broader in the \textit{Swift} light curve. One can now see a notable secondary peak in the \textit{Swift} folded light curve, which is separated by 0.5 in phase from the main peak. This second peak is also visible in the \textit{MAXI} light curve,  roughly 0.5 in phase after the primary, but it is not as clear as in the higher-energy \textit{Swift} light curve. There are also hints of other peaks in the folded light curves, particularly the \textit{MAXI} light curves, but it is hard to say whether these are above the error bars.

We also tried folding the \textit{Swift} and \textit{MAXI} light curves on the \textit{RXTE} period for a better comparison between the three data sets.  The light curves look very similar to those folded on the periods found for the different satellites. The consistency of the 34.5~d signal over close to 30 years leads very strong support to its interpretation as the orbital period of the system.

\section{Discussion}

We have carried out a detailed spectroscopic analysis of LS~437. In spite of the scarcity of diagnostic lines, we have obtained a consistent set of physical parameters for the star. The true spectral type of the star, if it were not distorted by the emission components, would be around O7.5\,Ve, with an absolute magnitude typical of the type, approaching $M_V=-5$. This makes LS~437 the earliest Oe star known in the Galaxy, comparable to the only other mid-Oe star, HD~155806 \citep{negueruela04_oe}. The mass of the star is poorly constrained, as the evolutionary mass $\approx28\:M_{\sun}$ is higher than the spectroscopic mass $\approx22\:M_{\sun}$. This discrepancy, however, is found for most O-type stars, and we can safely assume an intermediate value, $M_{*}\approx25\:M_{\sun}$, which is typical of the type \citep{martins05}.

The star shows an obvious underabundance of carbon, with the most likely value being 0.3~dex subsolar, but lower abundances being possible. Nitrogen is likely slightly enhanced with respect to the standard reference value. However, assuming single evolutionary tracks for single fast-rotating stars, a much higher N enhancement would be expected for a detectable C deficiency. Of course, all evolutionary models for the formation of Be/X-ray binaries imply that LS~437 underwent a phase of mass transfer from the progenitor of the neutron star, and this may help explain its unusual chemistry.

\subsection{Peculiar velocity}
Determining the radial velocity (RV) of LS~437 is not straightforward. There are few and very broad lines, and it is supposed to be a single-lined spectroscopic binary (SB1) system, where changes in RV reflect the orbital motion around the centre of mass. We measure the displacement of the line centres by fitting Gaussian profiles to the \ion{He}{ii} lines in all the high-resolution spectra. The measured radial velocity is around $110\: \mathrm{km\,s}^{-1}$ in the heliocentric reference system. The correction to the LSR for this direction is $17\: \mathrm{km\,s}^{-1}$ using the model of \citet{schonrich10}. Therefore, the radial velocity of LS~437, $v_{\mathrm{LSR}}$, should be between 90 and $95\: \mathrm{km\,s}^{-1}$. The model of \citet{Reid2014} predicts a radial velocity around $85\: \mathrm{km\,s}^{-1}$ at a distance of 7.6~kpc in this direction. Therefore, the radial component of the peculiar velocity is $\lesssim10\: \mathrm{km\,s}^{-1}$ for LS~437.

   \begin{figure}
   \centering
   \includegraphics[width=\columnwidth]{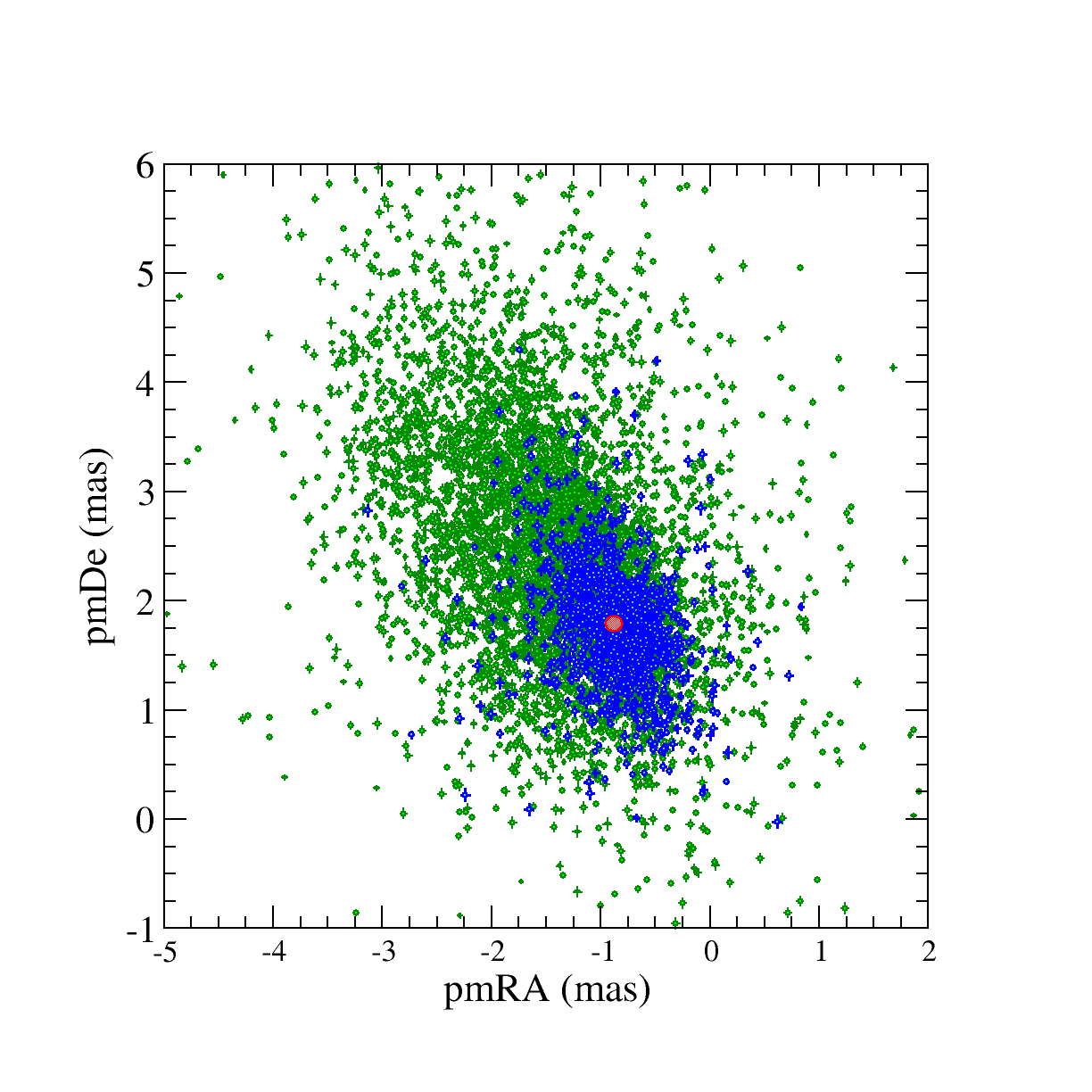}
      \caption{Vector point diagram for distant \textit{Gaia} sources within 1 degree of LS~437. The green points are objects with $\varpi<0.3$~mas, while the blue symbols mark $\varpi<0.1$~mas. The position of LS~437 is shown by the red circle. Its error bars are much smaller than the symbol. }
         \label{fig:vpd}
   \end{figure}

The tangential velocity is more difficult to estimate. In Fig.~\ref{fig:vpd}, we plot the vector point diagram (VPD) for distant objects within one degree of LS~437. Since we are probing very long distances and DR3 parallaxes may be negative, our only quality cut is an error in parallax smaller than $0.05$~mas. We select the sample of objects with $\varpi<0.3$~mas and mark (in blue) the subsample with $\varpi<0.1$~mas, i.e.\ distances comparable to that of LS~437. The position of LS~437 in the VPD coincides exactly with the highest density of objects in this direction, indicating that its proper motions are typical of distant stars in this part of the sky, which very likely suggests that they are dominated by Galactic rotation. Comparison with the field suggests that LS~437 is very unlikely to have a peculiar motion larger than 0.5~mas, corresponding to $<18\:\mathrm{km}\,\mathrm{s}^{-1}$ at 7.6~kpc.

As a further test, we search the recent catalogue of clusters found in \textit{Gaia} DR3 data by \citet{hunt24} for young open clusters with high distances in the neighbourhood. Even if we allow a separation in the sky of 5 degrees, there are only two such clusters, both with estimated distances around 6.5~kpc. In Table~\ref{tab:nearbyclusters}, we list their astrometric parameters and compare them to those of LS~437. Again, the differences are in both cases below 0.5~mas. 

In view of this, the 3D peculiar velocity of LS~437 is certain to be smaller than $20\:\mathrm{km}\,\mathrm{s}^{-1}$. Such a low value in a system that must have undergone a supernova explosion is not unexpected, given the high mass of the optical component. In general, BeX display $v_{\rm pec}\lesssim 40\:\mathrm{km}\,\mathrm{s}^{-1}$ \citep{nuchvani25}. The reason(s) for these low peculiar velocities is unclear, but they are usually thought to be a consequence of the evolutionary history leading to the system's formation. The progenitors of the neutron stars in BeX are generally believed to be reduced to small stellar cores with very thin envelopes in relatively wide orbits before the supernova explosion, resulting in weak natal kicks and low recoil velocities \citep[e.g.][and references therein]{larsen24,nuchvani25}.

\begin{table}
\caption{Astrometric parameters for distant young clusters in the vicinity of LS~437. Uncertainties represent standard errors for LS~437, but the standard deviation of cluster members for the clusters.\label{tab:nearbyclusters}}      
\centering
\begin{tabular}{l c c c}        
\hline\hline 
\noalign{\smallskip}
Object & $\varpi$ (mas) & pmRA & pmDE \\
& & (mas/yr) & (mas/yr)\\
\noalign{\smallskip}
\hline
\noalign{\smallskip}
LS~437 & $0.096\pm0.017$& $-0.88\pm0.01$ & $+1.79\pm0.02$  \\
HSC~1877 &$0.119\pm0.015$   & $-1.09\pm0.06$ & $+1.73\pm0.08$ \\
HSC~1914 &$0.120\pm0.02$ & $-0.80\pm0.05$ &  $+1.39\pm0.06$\\
\noalign{\smallskip}
\hline                        
\end{tabular}
\end{table}

\subsection{Rotational velocity}
Be stars are known to be fast rotators \citep{townsend04}. The critical velocity of a star with the observed parameters of LS~437 is in excess of $700\:\mathrm{km\,s}^{-1}$, but its measured projected rotational velocity is only $v~\sin\,i \approx 155\:\mathrm{km\,s}^{-1}$. Assuming a typical value $v_{\textrm{rot}}=0.8 \cdot v_{\textrm{crit}}$, the observed $v~\sin\,i$ would imply an inclination angle around $i=15\degree$. Such low inclination is at odds with the shape of emission lines, which should show strong wine-bottle-like inflections.

Motivated by this inconsistency, we rechecked the value of $v~\sin\,i$ by applying a somewhat different approach than used by \texttt{iacob-broad}. We applied the Fourier transform method, accepted the value given by the first zero and determined the macroturbulence to fit the profile, thus ignoring the GOF method, which is generally preferred by \texttt{iacob-broad}. For \ion{He}{i}~4387\,\AA, we find now $v\sin\,i = 170\: \mathrm{km\,s}^{-1}$, which is fully consistent with the \texttt{iacob-broad} value within the errors. An attempt to measure the velocity on the \ion{He}{i}~4026\AA\ line, which has a lower SNR, gives $v\sin\,i = 135\: \mathrm{km\,s}^{-1}$, with a higher $v_{\textrm{mac}}=195\: \mathrm{km\,s}^{-1}$. All the measurements from \ion{He}{i} lines give consistent results and put an upper limit on the total broadening (combined effect of rotation and macroturbulence) around $220\: \mathrm{km\,s}^{-1}$.

We then applied the technique to the \ion{He}{ii} lines, although the values obtained are not expected to be very accurate. It is well known that pressure broadening (such as the Stark broadening effect) may significantly affect H and He lines. When $v\sin\,i$ is not very high, \ion{He}{i} lines are preferable as rotational velocity diagnostics, because they are less affected by Stark broadening than \ion{He}{ii} lines \citep[see e.g.][]{simon06, ragudelo13, ssimon14}.
On the other hand, according to stellar surface models \citep{abdul23}, the width of \ion{He}{ii} lines in a fast-rotator seen at low inclination should be quite smaller than the width of \ion{He}{i} lines, as the \ion{He}{ii} lines form preferentially in the hotter polar regions. Nevertheless, we consistently find values around $v\sin\,i \approx 200\: \mathrm{km\,s}^{-1}$ for the \ion{He}{ii}, with a dispersion around 10\%. If we take into account the derived $v_{\textrm{mac}}$, the \ion{He}{ii} lines seem to be significantly broader than the \ion{He}{i} lines, although the $v\sin\,i$ derived are almost consistent within the errors. This is in agreement with the higher expected pressure broadening and strongly suggests that LS~437 is not a fast rotator seen at low inclination.

In any event, if we assume that the inclination angle is in the range $i=30\degree-40\degree$, as suggested by the shape of the emission lines\footnote{The results of \citet{sigut20} suggest that the inclination could be even higher, although none of the Be stars in their sample is as early as LS~437, which makes this uncertain.}, the only possible conclusion is that the star cannot be rotating at much more than half its critical speed.

\subsection{A persistent Be/X-ray binary?}

Recently, \citet{lapalombara25_0726} have suggested that X0726$-$260 must be grouped together with the low-luminosity persistent BeX, despite an average X-ray luminosity about one order of magnitude higher than typical of this group. Persistent low-luminosity BeX were first identified as a distinct subclass by \citet{reig1999} based on the absence of bright X-ray outbursts, persistent X-ray luminosity typically in the $L_{\mathrm{X}}\approx10^{34}$\,--\,$10^{35}\:\mathrm{erg}\,\mathrm{s}^{-1}$ range, and little variability. Other characteristics of the type include long pulse periods, very weak 6.4~keV iron lines in the X-ray spectra and possibly a soft thermal excess in the X-ray spectrum \citep{lapalombara25_class}. This set of characteristics may be interpreted in terms of systems with wide orbits and low eccentricities \citep{reig1999} in which the neutron star accretes directly from a low-density outflow without forming an accretion disk \citep[but see][for an alternative explanation involving a cold disk]{tsygankov17_cold}. These orbital characteristics suggest that a supernova explosion with a weak kick is necessary to form a persistent Be/X-ray binary.  

\citet{podsiadlowski04} argued that initial primaries with masses in the $8$\,--\,$11\:\mathrm{M}_{\sun}$ range may experience electron-capture supernovae, which will lead to small kick velocities. A second possibility is a very efficient second phase of mass transfer from the neutron star progenitor while it is a He star (case BB) that strips away its whole envelope and reduces it to a naked core, resulting in an ultra stripped supernova \citep{richardson23}. Case BB seems to be very frequent in BeX formation \citep[see, e.g.][]{rocha24}. The first scenario, which has recently been shown to probably apply to the persistent BeX RX~J0146.9+6121 \citep{marco25}, does not seem relevant for X0726$-$260, given the very high mass of LS~437. The second could only be of application if the peculiar velocity was very close to zero, which does not seem to be the case.

There is little reason to support the inclusion of X0726$-$260 in the group of persistent BeX. For a start, its X-ray luminosity is higher than those typically shown by these objects. Additionally, the short orbital period of X0726$-$260, fully within the range seen in transients with strong outbursts, sets it apart from the persistent BeX. The archetype of the class, X~Persei, has an orbital period of 250~d, and the orbital periods of most other persistent systems have not been determined, presumably because they are long. Even systems with relatively short spin periods, such as 1RXS~J225352.8+624354 ($P_{\mathrm{sp}}=46.7$~s), show evidence for quite longer orbital periods \citep{esposito13}.

%
   \begin{figure}
   \centering
   \includegraphics[width=\columnwidth]{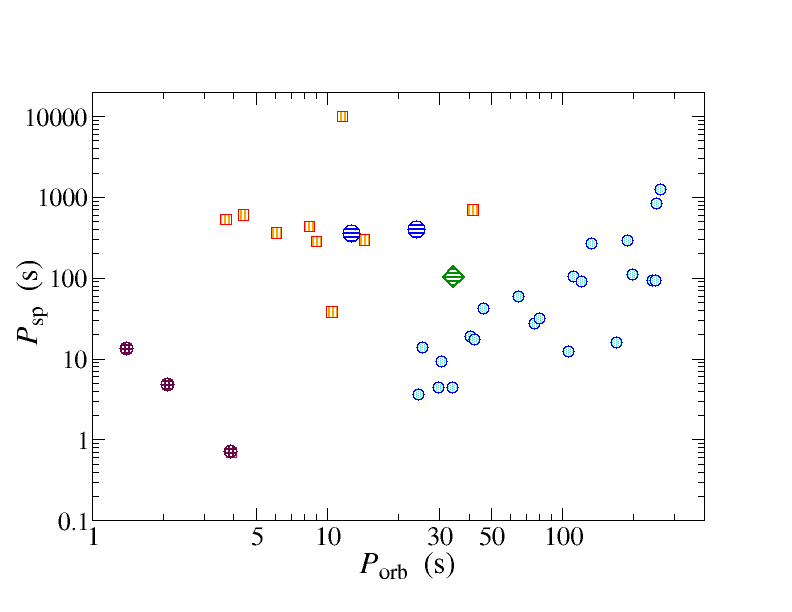}
      \caption{Corbet diagram for a large sample of HMXBs, taken from \citet{fortin23}. Blue circles represent BeX. Orange squares are (supergiant) wind accretors, while maroon circles are Roche-lobe overflow systems. The two BeX with anomalous positions, namely SAX~J2103.5+4545
 and 1A~1118$-$615 are represented by larger striped blue circles. The intermediate position of X~0726$-$260 is displayed by the green diamond.}
         \label{fig:corbet}
   \end{figure}

While the long-term lightcurves of persistent BeX are characterised by secular variations and occasional flaring \citep[e.g.][for X~Per]{lutovinov12}, the light curve of X0726$-$260 is dominated by strong orbital modulation. The \textit{Swift} lightcurve, corresponding to hard X-rays, shows very low luminosity at some orbital phases, a very strong peak that covers about one third of the orbital period, and a weaker peak in antiphase that covers about one tenth of the period. This strong modulation must arise from a non-negligible orbital eccentricity, again at odds with the inferred properties of persistent Be/X-ray binaries.

The orbital parameters of X0726$-$260 cannot be very dissimilar to those of the BeX transient V0332+53, which presents $P_{\mathrm{orb}}=33.9$~d and $e=0.37$ \citep{doroshenko16}. This source, which also contains an Oe star \citep{negueruela99}, is a transient, displaying occasional Type~II outbursts approaching $L_{{\textrm{X}}}\sim10^{38}\:\textrm{erg\,s}^{-1}$ and series of Type~I outbursts. As a matter of fact, the orbital characteristics of X0726$-$260 are fully typical of BeX transients.

\subsection{Origin of the X-rays}
\label{sec:origin}

The closest analogue to X0726$-$260 in terms of X-ray behaviour is the peculiar HMXB 4U~2206+54. This system consists of a neutron star orbiting the O9.5\,Vpe star BD~+53$^\circ$2790 \citep{negreig01}. It displays persistent X-ray emission, with typical luminosities around $\sim 10^{35}\:\textrm{erg\,s}^{-1}$, having sometimes reached $10^{36}\:\textrm{erg\,s}^{-1}$ \citep{blay05}. On short timescales, it is characterised by erratic flaring and substantial variability, leading to the suggestion that it is a wind-accreting object \citep{masetti04}. Two pointed observations of X0726$-$260, one with the \textit{RossiXTE}/PCA in 1997 \citep{corbet97} and the second with \textit{XMM-Newton} in 2023 \citep{lapalombara25_0726}, found substantial aperiodic variability, with evidence for flaring in the first observation. Contrarily, an observation with \textit{AstroSat} in 2016 found a very stable count rate, with variations below the 50\% level \citep{roy20}. Such behaviour is typical of wind-fed sources. 

Further evidence in this sense can be found in the $P_{\textrm{orb}}$/$P_{\textrm{sp}}$ (Corbet's) diagram (Fig.~\ref{fig:corbet}). In this plot, BeX display a loose correlation between the two values, which has been attributed to some form of quasi-equilibrium between spin-up during outbursts and spin-down during quiescence \citep[e.g.][]{waterskerkwijk,liu96,xu19}. Contrarily, all the wind-fed systems fall in an approximate straight line with spin periods of a few hundred seconds, with the exception of the peculiar system OAO~1657$-$415, which has a very evolved mass donor and is likely in a different evolutionary phase \citep{mason12} and the supergiant system 2S~0114+650, which has an extremely long (2.7~h) pulse period and is quickly spinning up \citep{farrell08}. X0726$-$260 occupies a strange position in this diagram, well above the BeX, but below the wind-fed systems, a position that can be considered intermediate between the two groups.

The early spectral type of LS~437 makes the possibility of a wind-fed system more likely. Using the LIME simulator \citep{sundqvist25}, we find that the wind mass-loss rate for a star like LS~437 is on the order of $\dot{M} \approx 1.6\times10^{-7}\:\mathrm{M}_{\sun}$, at least four times higher than for a more typical O9.5\,Ve counterpart, suggesting that accretion from the stellar wind may contribute rather more significantly than in any other BeX known. Of course, a key parameter determining the efficiency of wind accretion is the wind terminal velocity \citep[see, e.g.][]{hainich20}, which we cannot directly measure with existing data. In the case of 4U~2206+54, both \textit{IUE} \citep{ribo06} and \textit{HST} \citep{hainich20} observations reveal a very slow wind ($v_{\infty}\approx400\:\textrm{km\,s}^{-1}$), which allows a significantly increased accretion rate with respect to the tabulated value for the spectral type \citep{ribo06}. 

Nevertheless, some simple geometrical considerations and the application of Kepler's laws allow a meaningful comparison, which is shown in Table~\ref{tab:winds}. Barring the unknown wind speed in LS~437, we can see a quite comparable orbital configuration in both systems, while the mass loss rate in LS~437 is expected to be about 5 times higher. It is worth noting that 4U~2206+54 displayed, during the first few years of \textit{RXTE}/ASM monitoring, a strong modulation at $\sim9.56$ days \citep{corbet01}, which was assumed to be the orbital period. After some time, the dominant peak became the frequency corresponding to $\sim 19.25$ days, almost exactly twice this value \citep{corbet07}. With the new ephemeris, the source displays weak X-ray emission peaks half a phase away from the main peak, suggesting that apastron luminosity was initially almost as high as that near periastron, leading to the first periodicity detected. In the case of X0726$-$260, the observed periodicity is almost certainly the orbital period, as there seems to be a secondary increase in X-ray luminosity close to half a period after the main peak, but varying in phase.

Another wind-fed system with moderately increased emission at a phase not corresponding to the X-ray peak is IGR~J00370+6122, whose counterpart is a low-luminosity BN0.7\,Ib supergiant \citep{gonzalez14}. In this system, the mass donor is much larger than LS~437, with an estimated $R_{*}\simeq17\:\mathrm{R}_{\sun}$ \citep{gonzalez14,hainich20}, and the orbit has a high eccentricity $e=0.56$, allowing the neutron star to come close to the surface of the supergiant (see Table~\ref{tab:winds}). This led \citet{hainich20} to speculate that the X-ray peaks, which happen some time after periastron, might be related to localised Roche-lobe overflow, although this is probably not necessary to explain the moderate increase in X-ray luminosity. Away from the peak, the luminosity of this system is somewhat below $10^{35}\:\textrm{erg\,s}^{-1}$.  The mass loss rate from the supergiant is similar to that of BD~+53$^\circ$2790, but the terminal velocity in this case is a rather more typical 1100~km\,s$^{-1}$. Again, consideration of the orbital geometry close to apastron shows that the wind parameters of LS~437 do not need to be terribly unusual to produce a comparable X-ray luminosity.

\begin{table}
\caption{Properties of some peculiar wind-fed system with moderate X-ray luminosity.\label{tab:winds}}      
\centering
\begin{tabular}{l c c c c}        
\hline\hline 
\noalign{\smallskip}
Object & $M_{*}$  & $\dot{M}_{*}$  &$a$ & $e$ \\
&($\mathrm{M}_{\sun}$) & ($10^{-8}\cdot\mathrm{M}_{\sun}$)& ($\mathrm{R}_{*}$) &\\
\noalign{\smallskip}
\hline
\noalign{\smallskip}
X0726$-$260 & 25& $16$ &13.8 &0.2$^{1}$ \\
4U~2206+54$^{2}$ & 16   & 3.2 & 10.7 & 0.15 \\
IGR~J00370+6122$^{3}$ &15 & 3.2 &  4 & 0.56 \\
\noalign{\smallskip}
\hline                        
\end{tabular}
\begin{minipage}{\textwidth} 
\vspace{0.25cm}
\footnotesize
Notes:\\
$^1$ Assumed.\\
$^2$ Data from \citet{ribo06}, except for $\dot{M}_{*}$, from \citet{hainich20}.\\
$^3$ Data from \citet{gonzalez14}, except for $\dot{M}_{*}$, \\from \citet{hainich20}.\\
\end{minipage}
\end{table}

\subsection{Disk size and stability}

Although the X-ray luminosity of X0726$-$260 can be explained as a consequence of wind accretion in a moderately eccentric orbit, the observed spectral features of LS~437 can only be explained by the presence of a disk around the primary\footnote{Following convention, LS~437 is referred to as the primary in the current binary, because it is by far the most massive component. The progenitor of the neutron star, which was the more massive component before mass transfer, is generally termed the initial or original primary.}. In fact, the main difficulty in explaining the X-ray behaviour of the system is understanding the stability of the X-ray emission and the complete absence of reported outbursts in about fifty years of observations. Well studied BeX transients, such as 4U\;0115+63 or A\;0525+262, display cycles of variability, during which features associated with their disks change considerably, sometimes disappearing completely. Along these cycles, they experience Type~I and Type~II outbursts \citep{negueruela2001,reig2005}. Systems with close ($P_{\textrm{orb}}\lesssim100$~d) orbits and moderate eccentricities are characterised by series of Type~I outbursts \citep{okaneg01}.
Type~II outbursts are frequently associated with major changes in the emission spectrum, leading to the suggestion that large-scale perturbations of the disks lead to their onset \citep[e.g.][]{negueruela2001, martin14}. The behaviour of X0726$-$260 has been markedly different. 

Among BeX, V0332+53 is the most similar to X0726$-$260 in terms of early primary ($\approx\:$O8.5\,Ve) and orbital period (33.9~d). This system is less active than other short-period, moderate-eccentricity systems, having undergone three major outbursts during the past fifty years. The high temperature of the primary is expected to be reflected in a higher disk temperature, which in turn results in a higher viscous torque \citep{okaneg01}. The higher viscous torque makes truncation by the neutron star more difficult, but also leads to a more stable disk. On the other hand, the high mass of the counterpart results in a larger mass ratio, which results in more efficient tidal truncation \citep{martin24}. \citet{franchini19} studied the mechanisms that could give rise to Type~I outbursts in low-eccentricity systems, such as XTE~1948+32 or 2S~1553$-$542, concluding that they might not work for Be disks that present a large misalignment with the orbital plane or are flared at the truncation radius. At moderate eccentricities, misaligned disks lead to Type~I outbursts and will likely evolve quickly to produce Type~II outbursts, though \citep{martin14}. 

None of these scenarios seem to apply to LS~437.
Although the high-resolution spectra suggest some variation in the relative strength of the blue and red peaks of the emission lines, the features are surprisingly stable over a very long time span. Moreover, the disk size measured with Huang's approximation (Table~\ref{tab:disksize}) suggests for all the H and He lines a disk larger than the expected periastron distance (Table~\ref{tab:winds}). Only the metallic lines are barely compatible with this size. This suggests a breakdown in the assumptions implicit in the approximation, and a departure from the typical properties of Be disks.

\section{Conclusions}

We have carried out a detailed spectroscopic study of LS~437, the optical component of X0726$-$260. We derive a moderate projected rotational velocity $v\,\sin\,i\approx 160\:\mathrm{km\,s}^{-1}$, which would necessitate a very low inclination angle if the star was rotating at a significant fraction of its critical velocity. Such a low inclination is incompatible with the observed morphology of the emission lines, which instead indicate a moderate viewing angle. From our analysis, we obtain $T_{\textrm{eff}}\approx36$~kK adopting $\log\,g\approx3.8$, in good accord with an apparent spectral type O7.5\,V. Such early type favours the possibility that the X-ray emission arises, at least in part, from wind accretion. In fact, the position of X0726$-$260 in the Corbet diagram and the shape of its X-ray lightcurve suggest that both residual accretion from a truncated disk and wind accretion may contribute.

The system is primarily  characterised by its long-term stability, both in  X-ray luminosity and the morphology of emission features. There are no reported X-ray outbursts in about fifty years of monitoring, with luminosity variations remaining within an order of magnitude. Similarly, the source has always displayed moderately strong emission lines (EW of H$\alpha$ between $-6$ and $-10$\,\AA), with little morphological change.  Such behaviour could be easily explained by effective disk truncation by the neutron star companion in a low-eccentricity orbit. However, the strong modulation of the X-ray emission with a 35.5 d period, which must be interpreted as the orbital period of the system, prevents a negligible eccentricity. The BeX transient with more similar orbital parameters to X0726$-$260, V0332+53, has a measured $e=0.37$ and displays Type~II outbursts. The very different X-ray behaviours of the two sources are reflected in their spin periods, 4.4~s for V0332+53 and 103~s for X0726$-$260. If we attribute these diverse behaviours to the eccentricity of the systems, then the combined constraints of a measurable $e$ to display orbital modulation and a significantly lower value than in V\,0332+53, suggest that the eccentricity in X0726$-$260 cannot be far from 0.2. As this is a fundamental parameter to constrain the origin of the X-ray emission, we will attempt a full orbital solution using a much expanded spectroscopic dataset in a future paper. 

In all, the unusual properties of X0726$-$260, whose closest analogue in terms of X-ray behaviour may be considered 4U\,2206+54, together with its very hot and massive main sequence companion, add to the growing complexity of high-mass X-ray binaries. As we accumulate more information about their long-term behaviour, it becomes increasingly difficult to neatly separate these systems into discrete categories.
   
\begin{acknowledgements}

We thank Sergio Sim\'on-D\'{\i}az for access to non-public spectra of Oe stars within the IACOB database and Francesc Vilardell for the reduction of the EMMI spectrum. We also thank the anonymous referee for detailed reading and useful suggestions.\\
This research is partially supported by the Spanish
Government Ministerio de Ciencia, Innovaci\'on y Universidades and Agencia Estatal de Investigación (MCIU/AEI/10.130 39/501 100 011 033/FEDER, UE) under
grants PID2021-122397NB-C21/C22 and PID2024-159329NB-C21/22. It is also supported by MCIU with funding from the European Union NextGenerationEU
and Generalitat Valenciana in the call Programa de Planes Complementarios
de I+D+i (PRTR 2022), project HIAMAS, reference ASFAE/2022/017. SRB acknowledges the support of the Viera y Clavijo program funded by the Agencia Canaria de Investigación, Innovación y Sociedad de la Información del Gobierno de Canarias and the Universidad de la Laguna. LJT acknowledges support from the South African National Research Foundation. Some of the data presented in this paper were obtained with the Southern African Large Telescope under the proposals 2019-2-MLT-004, 2021-2-MLT-005 \& 2023-2-MLT-005 (PI: Townsend). \\
This research has made use of the Simbad database, operated at CDS,
Strasbourg (France). 
This work has made use of data from the European Space Agency (ESA) mission Gaia
(https://www.cosmos.esa.int/gaia), processed by the Gaia Data Pro-
cessing and Analysis Consortium (DPAC, https://www.cosmos.esa.int/
web/gaia/dpac/consortium). Funding for the DPAC has been provided by
national institutions, in particular the institutions participating in the Gaia Multi-lateral Agreement. 
\end{acknowledgements}

%
%

\bibliographystyle{aa} 
\bibliography{bins,bes,obstars,gaia,class,instro}
  

\begin{appendix}

\section{Fitting line models} \label{ApA}
\begin{figure*}[ht]
   \centering
   \includegraphics[width=17cm, angle=180,trim={0cm 1cm 0cm 4.5cm},clip]{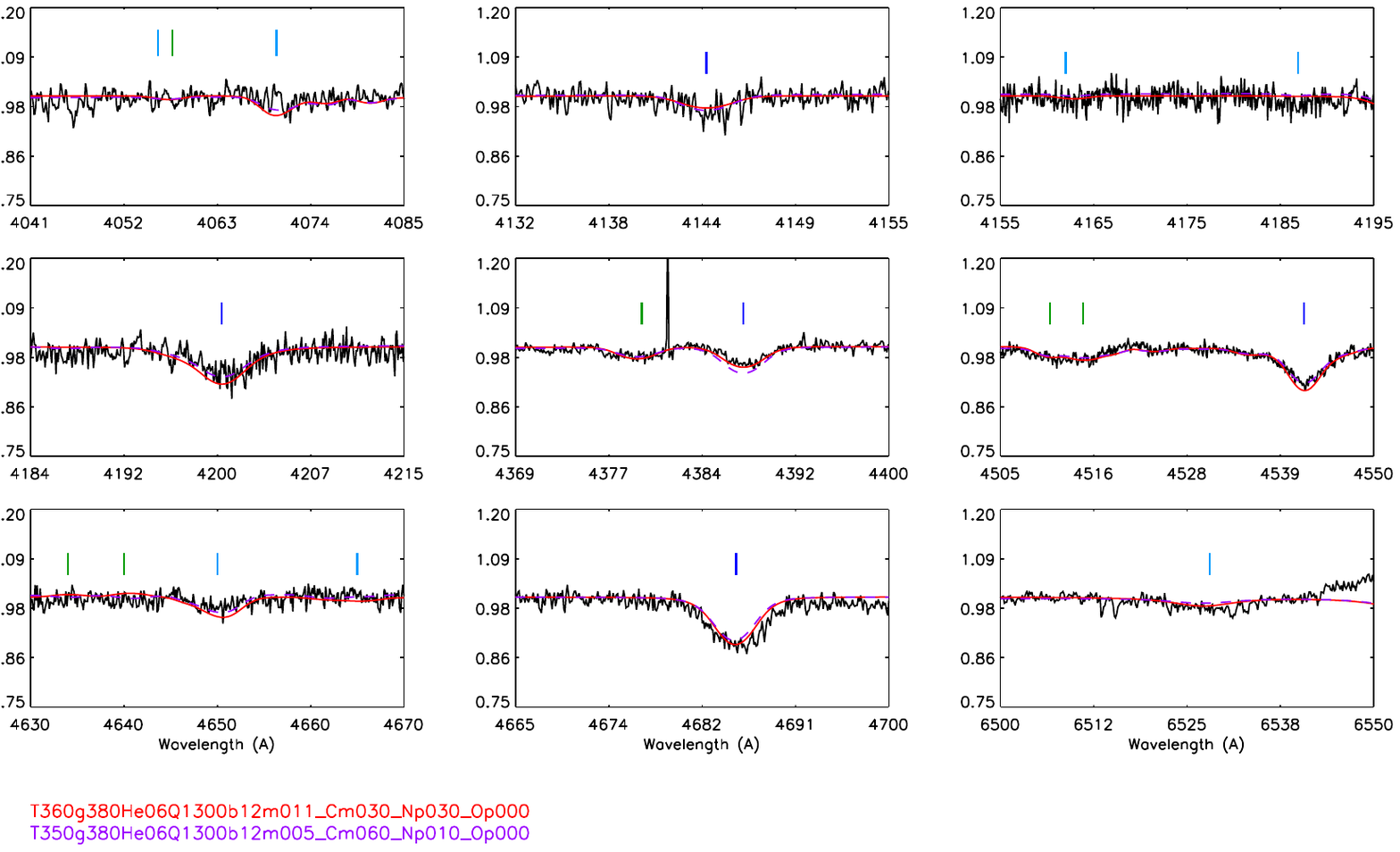}
      \caption{Best-fit HHeCN models to the observed MEGARA spectrum of LS~437. With a red solid line is shown the model with parameters:  $T_{\mathrm{eff}}$  = 36.0 kK, $\log\,g$  = 3.8 dex, $\xi$ = 11.0 km s$^{-1}$, $Y$(He) = 0.06,  log (C/H) = 8.05 dex, log (N/H) = 8.10 dex. With a purple dashed line the model with parameters:  $T_{\mathrm{eff}}$  = 35.0 kK, $\log\,g$   = 3.8 dex, $\xi$ = 5.0 km s$^{-1}$, $Y$(He) = 0.06,  log (C/H) = 7.75 dex, log(N/H) = 7.90 dex. C and N  lines used in the analysis are indicated with light blue and green vertical lines, respectively. He lines are indicated with dark blue vertical lines for reference. The abundance fits were visually validated against a sensitivity margin of $\pm0.05$~dex.}
         \label{bfm}
   \end{figure*}

Figure~\ref{bfm} shows the He, C, and N spectral lines used in our analysis and the FASTWIND best fitting models to the observed MEGARA spectrum of LS~437.

\end{appendix}

\end{document}